\documentclass[conference]{IEEEtran}
\IEEEoverridecommandlockouts
\usepackage{cite}
\usepackage{amsmath,amssymb,amsfonts}

\usepackage{subfigure}
\usepackage{algorithm}
\usepackage[noend]{algpseudocode}
\usepackage{algorithmicx}
\usepackage{listings}
\usepackage{url}
\usepackage{tabularx}
\usepackage{booktabs}
\usepackage{graphicx}
\usepackage{textcomp}
\usepackage{xcolor}
\usepackage[T1]{fontenc}
\usepackage[utf8]{inputenc}

\providecommand{\subjectto}{\ensuremath{\text{subject to}}}

\renewcommand{\vec}[1]{\textbf{#1}}

\providecommand{\mat}[1]{\boldsymbol{{#1}}}%

\providecommand{\mA}{\ensuremath{\mat{A}}}
\providecommand{\mB}{\ensuremath{\mat{B}}}
\providecommand{\mC}{\ensuremath{\mat{C}}}
\providecommand{\mD}{\ensuremath{\mat{D}}}

\providecommand{\mW}{\ensuremath{\mat{W}}}
\providecommand{\mX}{\ensuremath{\mat{X}}}

\providecommand{\va}{\ensuremath{\vec{a}}}
\providecommand{\vb}{\ensuremath{\vec{b}}}
\providecommand{\vc}{\ensuremath{\vec{c}}}

\providecommand{\vx}{\ensuremath{\vec{x}}}
\providecommand{\vy}{\ensuremath{\vec{y}}}

\def\BibTeX{{\rm B\kern-.05em{\sc i\kern-.025em b}\kern-.08em
    T\kern-.1667em\lower.7ex\hbox{E}\kern-.125emX}}
\begin{document}

\title{A Parallel Projection Method for Metric Constrained Optimization\\
\thanks{David Gleich is supported by the DARPA Simplex Program, the Sloan Foundation, and NSF awards CCF-1149756,  IIS-154648, and CCF-093937. Nate Veldt is funded by NSF award IIS-154648.}
}

\author{\IEEEauthorblockN{Cameron Ruggles}
\IEEEauthorblockA{\textit{Department of Computer Science} \\
\textit{Purdue University}\\
West Lafayette, IN\\
cruggles@purdue.edu}
\and
\IEEEauthorblockN{Nate Veldt}
\IEEEauthorblockA{\textit{Department of Mathematics} \\
\textit{Purdue University}\\
West Lafayette, IN\\
lveldt@purdue.edu}
\and
\IEEEauthorblockN{David F. Gleich}
\IEEEauthorblockA{\textit{Department of Computer Science} \\
	\textit{Purdue University}\\
	West Lafayette, IN\\
	dgleich@purdue.edu}
}

\maketitle

\begin{abstract}
	Many clustering applications in machine learning and data mining rely on solving \emph{metric-constrained} optimization problems. These problems are characterized by $O(n^3)$ constraints that enforce triangle inequalities on distance variables associated with $n$ objects in a large dataset. Despite its usefulness, metric-constrained optimization is challenging in practice due to the cubic number of constraints and the high-memory requirements of standard optimization software. Recent work has shown that iterative projection methods are able to solve metric-constrained optimization problems on a much larger scale than was previously possible, thanks to their comparatively low memory requirement. However, the major limitation of projection methods is their slow convergence rate. In this paper we present a parallel projection method for metric-constrained optimization which allows us to speed up the convergence rate in practice. The key to our approach is a new parallel execution schedule that allows us to perform projections at multiple metric constraints simultaneously without any conflicts or locking of variables. We illustrate the effectiveness of this execution schedule by implementing and testing a parallel projection method for solving the metric-constrained linear programming relaxation of correlation clustering. We show numerous experimental results on problems involving up to 2.9 trillion constraints.
\end{abstract}


\section{Introduction}
Many tasks in machine learning and data mining, in particular problems related to clustering, rely on learning pairwise distance scores between objects in a dataset of $n$ objects. One particular paradigm for learning distances, that arises in a number of different contexts, is to set up a convex optimization problem involving $O(n^2)$ distance variables and $O(n^3)$ \emph{metric constraints} which enforce triangle inequalities on the variables. This approach has been applied to problems in sensor location~\cite{gentile2005sensor,gentile2007distributed}, metric learning~\cite{biswas2014semi,batra2008semi}, metric nearness~\cite{brickell2008metricnearness,dhillon2003MNreport,dhillon2004tfa}, and joint clustering of image segmentations~\cite{glasner2011contour,Vitaladevuni2010coclustering}. Metric-constrained optimization problems also frequently arise as convex relaxations of NP-hard graph clustering objectives. A common approach to developing approximation algorithms for these clustering objectives is to first solve a convex relaxation and then round the solution to produce a provably good output clustering~\cite{chawla2015near,leighton1999multicommodity,veldt2017lamcc}.

The constraint set of metric-constrained optimization problems may differ slightly depending on the application. However, the common factor among all of these problems is that they involve a cubic number of constraints of the form $x_{ij} \leq x_{ik} + x_{jk}$ where $(i,j,k)$ is a triplet of points in some dataset and $x_{ij}$ is a distance score between two objects $i$ and $j$. This leads to an extremely large, yet very sparse and carefully structured constraint matrix. Given the size of this constraint matrix and the corresponding memory requirement, it is often not possible to solve these problems on anything but very small datasets when using standard optimization software. In recent work~\cite{veldt2018projection} we showed how to overcome the memory bottleneck by applying memory-efficient iterative projection methods, which provide a way to solve these problems on a much larger scale than was previously possible. Unfortunately, although projection methods come with a significantly decreased memory footprint, they are also known to exhibit very slow convergence rates. In particular, the best known results are obtained by specifically applying Dykstra's projection method~\cite{dykstra1983algorithm}, which is known to have a only a linear convergence rate~\cite{escalante2011altproj}. 

Given the slow convergence rate of Dykstra's method, a natural question to ask is whether one can improve its performance using parallelism. There does in fact already exist a parallel version of Dykstra's method~\cite{iusem1991convergence}, which performs independent projections at all constraints of a problem simultaneously, and then averages the results to obtain the next iterate. However, this procedure is ineffective for metric-constrained optimization, since averaging over the extremely large constraint set leads to changes that are so small no meaningful progress is made from one iteration to the next. As another challenge, we note that many of the most commonly studied metric-constrained optimization problems are linear programs~\cite{Agarwal2008metricmod,chawla2015near,gentile2005sensor,glasner2011contour,Vitaladevuni2010coclustering,veldt2017lamcc}. Because linear programming is P-complete, parallelizing LP solvers is in general very hard. Thus, finding meaningful ways to solve metric-constrained optimization problems in a way that is both fast and memory efficient possess several significant challenges.

In this work we take a first step in parallelizing projection methods for metric-constrained optimization. This leads to a modest but consistent reduction in running time for solving these challenging problems on a large scale. Our approach relies on the observation that when applying projection methods to metric-constrained optimization, two projection steps can be performed simultaneously and without conflict as long as the $(i,j,k)$ triplets associated with different metric constraints share at most one index in common. Based on this, we develop a new parallel execution schedule which identifies large blocks of metric constraints that can be visited in parallel without locking variables or performing conflicting projection steps. Because Dykstra's projection methods also relies on carefully updating dual variables after each projection, we also show how to keep track of dual variables in parallel and update them at each pass through the constraint set. We demonstrate the performance of our new approach by using it to solve the linear programming relaxation of correlation clustering~\cite{Bansal2004correlation}. Solving this LP is an important first step in many theoretical approximation algorithms for correlation clustering~\cite{AilonCharikarNewman2008,charikar2005clustering,chawla2015near,puleo2015cc,puleo2016cc,veldt2017lamcc}. In our experiments we consistently obtain a speedup of roughly a factor 5 over the serial method using even a small number cores, and achieve a speedup of over a factor of 11 for our largest problem. Our new approach allows us to handle problems containing up to nearly 3 trillion constraints in a fraction of the time it takes the serial method.
 
\section{Background}
We use the term \emph{metric-constrained optimization} or more simply \emph{metric optimization} to refer to any convex optimization problem involving constraints of the form $x_{ij} \leq x_{ik} + x_{jk}$ where $x_{ij}$ represents a distance variable between two points $i$ and $j$ in a large graph or dataset. Our work builds directly on previous results for solving optimization problems of this form using projection methods~\cite{sra2005MNnips,brickell2008metricnearness,veldt2018projection}. In this section we specifically consider the metric-constrained linear programming relaxation for correlation clustering and its relationship to what is known as the metric nearness problem. We will use this LP relaxation as a special case study in this paper, although the parallel approach we develop can in principle be applied to any metric optimization problem. 

\subsection{Metric Nearness and Correlation Clustering}
One key example of metric optimization is the metric nearness problem~\cite{sra2005MNnips,brickell2008metricnearness}, in which one is given matrix $\mD = (d_{ij})$ of dissimilarity scores between objects in a dataset. The goal is to find the matrix $\mX = (x_{ij})$ whose entries satisfy the triangle inequality and for some value of $p$ minimizes
\begin{equation}
\label{metricnear}
||\mX - \mD ||_p = \bigg( \sum_{ij} w_{ij}|x_{ij} - d_{ij}|^p \bigg)^{1/p},
\end{equation} 
where $w_{ij}$ is a nonnegative weight indicating the how strongly we wish $x_{ij}$ to be similar to $d_{ij}$. The problem can be cast as a linear program when $p = 1$, a quadratic program when $p = 2$, and a slightly more complicated convex optimization problem for other finite values of $p$. One can also consider a $p = \infty$ norm version of the problem which minimizes the the maximum of $|x_{ij} - d_{ij}|$ over all pairs $i,j$. This can also be cast as an LP.

Metric-constrained optimization is also a key ingredient in approximation algorithms for correlation clustering~\cite{Bansal2004correlation}. In correlation clustering one is given a weighted and signed graph $G = (V, E^+, E^-, W)$. Each pair of nodes $(i,j)$ in $G$ defines either a positive edges $(i,j)\in E^+$ or a negative edges $(i,j) \in E^-$. The goal is to partition $V$ in such a way that negative edges tend to link nodes between different clusters, and positive edges link nodes inside the same cluster. The problem also comes with weights $W = (w_{ij})$ where $w_{ij}$ indicates the strength of the relationship between $i$ and $j$. One formulation of the problem is to minimize the weight of mistakes, which can be cast as the following binary linear program:
\begin{equation}
\label{cc}
\begin{array}{lll} \text{minimize} & \sum_{(i,j)\in E^+} w_{ij}x_{ij} \,\,+ &\sum_{(i,j) \in E^-} w_{ij}(1-x_{ij}) \\ \subjectto & x_{ij} \leq x_{ik} + x_{jk} &  \text{ for all $i,j,k$} \\ & x_{ij} \in \{0,1\} &\text{ for all $i,j$}. \end{array}
\end{equation}
A positive mistake happens when two nodes with a positive edge are clustered apart ($x_{ij} = 1$), and this comes with a penalty equal to the weight $w_{ij}$. A negative mistake is when two nodes sharing a negative edge are clustered together, in which case the penalty is again $w_{ij} = w_{ij}(1-x_{ij})$ since in this case $x_{ij} =1$. We can relax~\eqref{cc} to a linear program by substituting $x_{ij} \in \{0,1\}$ with the constraint $x_{ij} \in [0,1]$. Solving this relaxation and then rounding the solution is a general strategy that has lead to a number of approximation algorithms for different variants of correlation clustering. For arbitrary weights, there exists an $O(\log n)$ approximation rounding scheme~\cite{DemaineEmanuelFiatEtAl2006}. When the graph is unweighted (i.e. $w_{ij} = 1$ for all pairs $i,j$), the best rounding scheme produces an approximation ratio near 2~\cite{chawla2015near}. Several other special weighted cases also obtain their best known approximation factor by solving the relaxation of~\eqref{cc} and rounding~\cite{veldt2017lamcc,puleo2016cc,AilonCharikarNewman2008}.

In recent work~\cite{veldt2018projection} we proved that the LP relaxation of~\eqref{cc} can be cast equivalently as a special case of the metric nearness problem~\eqref{metricnear} when $p = 1$. Specifically, given an instance of correlation clustering, define a dissimilarity score $d_{ij} = 1$ if $(i,j) \in E^-$, and set $d_{ij} = 0$ otherwise. Then the $\ell_1$ metric nearness problem and the LP relaxation of correlation clustering are both equivalent to the following metric-constrained LP:
\begin{equation}
\label{mn1}
\begin{array}{lll} \text{minimize} & \sum_{i < j} w_{ij} f_{ij} & \\ \subjectto  &  x_{ij} \leq x_{ik} + x_{jk} & \text{ for all $i,j,k$} \\ & x_{ij} - d_{ij} \leq f_{ij} & \text{ for all $i,j$} \\ & d_{ij}-x_{ij} \leq f_{ij} &\text{ for all $i,j$}. \end{array}
\end{equation}

\subsection{Projection Methods for Metric Optimization}
Although solutions to problems such as~\eqref{metricnear}, \eqref{mn1}, and other metric optimization problems are desirable from a theoretical perspective, they are challenging to solve for even modest values of $n$ due to the cubic constraint set. Standard commercial optimization packages are typically unable to handle problems with even a few hundred nodes when the full constraint set is included, due to memory limitations. Sra et al. began to address this problem specifically for the metric nearness problem~\cite{sra2005MNnips}. Their approach was to apply memory-efficient projection methods, which visit constraints cyclically and iteratively update variables in a manner that is proven to converge to the optimal solution. 

Recently, we showed how the techniques of Sra et al. can be adapted and improved to apply more broadly to a wider range of linear and quadratic metric-constrained optimization problems~\cite{veldt2018projection}. These results come with new approximation guarantees for specific graph clustering objectives, and are designed to produce output solutions with better constraint satisfaction and convergence guarantees. Here we review the main background for applying Dykstra's method to metric-constrained linear programming. For details on how to apply projection methods to metric-constrained convex optimization problems that are not linear programs, we refer to the reader to other work~\cite{brickell2008metricnearness,sra2005MNnips}. 


\paragraph{Metric-Constrained Linear Programming}
Consider a general linear program of the form 
\begin{equation}
\label{lp}
\min \,\, \vc^T \vx \text{  s.t. } \mA \vx \leq \vb.
\end{equation}
Encoding a metric-constrained LP in this format can be accomplished by letting $\vx$ encode a linearization of the distance variables $x_{ij}$ and potentially other variables depending on the specific optimization problem. The constraint matrix $\mA$ will encode metric constraints and other problem specific constraints, (e.g. the non-metric constraints $x_{ij} - d_{ij} \leq f_{ij}$ in \eqref{mn1}). Because of the metric constraints, $\mA$ will be large, sparse, and very structured. 

Projection methods do not apply directly to solving linear programs, so we first consider a regularized linear program
\begin{equation}
\label{qp}
\min \vc^T\vx + \frac{\varepsilon}{2} \vx^T \mW \vx \text{  s.t. } \mA \vx \leq \vb
\end{equation}
where $\varepsilon$ is a positive constant and $\mW$ is a positive definite diagonal matrix of weights. Both $\varepsilon$ and $\mW$ are viewed as parameters that can be chosen to control the relationship between~\eqref{lp} and~\eqref{qp}. When $\mW$ is the identity matrix, solving~\eqref{qp} for a small enough value of $\varepsilon$ will output the smallest norm solution to the LP~\eqref{lp}~\cite{mangasarian1984normal}. Furthermore, our recent work provides specific details for how to set $\varepsilon$ and $\mW$ to bound the difference between the original linear program and the related quadratic program~\eqref{qp} for specific graph clustering relaxations~\cite{veldt2018projection}. 


\paragraph{Applying Projection Methods}
The quadratic program~\eqref{qp} can be solved using memory-efficient projection methods, which iteratively visit constraints and perform correction and projection steps that slowly fix constraint violations, update dual variables, and eventually converge to the unique optimal solution. Following previous work~\cite{sra2005MNnips,veldt2018projection}, we specifically consider Dykstra's method, which for quadratic programs is equivalent to Hildreth's method~\cite{hildreth1957quadratic} and Han's method~\cite{han1988successive}. We provide pseudocode for applying this method to~\eqref{qp} in Algorithm~\ref{alg:dykstraqp}. 
\begin{algorithm}[tb]
	\caption{Dykstra's Method for Quadratic Program~\eqref{qp}}
	\begin{algorithmic}[5]
		\State $(M,N)$ = number of rows and columns of $\mA$ respectively
		\State $\vy := \textbf{0} \in \mathbb{R}^M$ (dual variables)
		\State $\vx := -\frac{1}{\varepsilon}\mW^{-1}\vc$, $k := 0$ 
		\While{\emph{not converged}}
		\State $k := k+1$
		\State (Visit next constraint): $i := (k-1) \bmod M + 1$ 
		\State (Correction step):  $\vx := \vx + y_i (\frac{1}{\varepsilon}\mW^{-1} \va_i)$ 
		\State \hspace{1cm} where $\va_i$ is the $i$th row of $\mA$
		\State (Projection step): $\vx := \vx - \theta_i^+ (\frac{1}{\varepsilon}\mW^{-1} \va_i)$
			\State \hspace{1cm}  where $\theta_i^+ = \varepsilon \frac{ \max \{\va_i^T \vx - b_i, 0 \}}{\va_i^T \mW^{-1} \va_i }$
		\State (Dual variable update): $y_i := \theta_i^+ \geq 0$
		\EndWhile
	\end{algorithmic}
	\label{alg:dykstraqp}
\end{algorithm}

\paragraph{Localized Metric Projections}
For a more in-depth explanation of the algorithm, we refer to our previous work~\cite{veldt2018projection}. The key thing to realize is that Algorithm~\ref{alg:dykstraqp} is simply Dykstra's method applied specifically to solve~\eqref{qp}. Most importantly, updates of the form $\vx := \vx + c\mW^{-1}\va_i$ for a constant $c$ can be performed very quickly for metric constraints, since in this case $\va_i$ (the $i$th row of constraint matrix $\mA$) has only three nonzero entries.

For illustration, we show how to perform the projection step in Algorithm~\ref{alg:dykstraqp} when $\vx = (x_{ij})$ is a linearization of the distance variables, and $\mW$ is the identity matrix (note that the projection step is unaffected by the value of $\varepsilon$, so we do not specify its value here). Row $\va$ of $\mA$ and entry $b$ of $\vb$ encode the constraint $\va^T \vx = x_{ij} - x_{ik} - x_{jk} \leq 0 = b$. For this constraint, $\va$ has three nonzero entries: 1, $-1$, and $-1$, corresponding to the locations of $x_{ij}$, $x_{ik}$, and $x_{jk}$ in $\vx$.  If $\delta = \va^T\vx = x_{ij} - x_{ik} - x_{jk} \leq 0$, then the constraint is already satisfied, and $\max \{\va^T \vx - b, 0 \} = 0$, thus there is no update to the vector $\vx$. If $\delta > 0$, then $ \max \{\va^T \vx - b, 0 \}  = \delta = x_{ij} - x_{ik} - x_{jk}$, and $\va^T \va = 3$. The projection step in Algorithm~\ref{alg:dykstraqp} updates only three entries of $\vx$:
\[ x_{ij} \leftarrow x_{ij} - \delta/3, \hspace{.5cm} x_{ik} \leftarrow x_{ik} +\delta/3, \hspace{.5cm} x_{jk} \leftarrow x_{jk} +\delta/3. \]
The correction step in Algorithm~\ref{alg:dykstraqp}, which is necessary to guarantee convergence, can be performed in a similar localized manner.

\paragraph{Slow Convergence Rate}
The decreased memory footprint of Dykstra's method makes it possible to solve metric constrained problems on a much larger scale than was previously possible~\cite{veldt2018projection}. However, this method converges very slowly, given that the convergence rate for Dykstra's method applied to quadratic programs is only linear~\cite{escalante2011altproj}. 

\section{Parallel Metric Constrained Optimization}
The primary contribution of our work is to show how to parallelize projection methods specifically for metric-constrained optimization problems. We accomplish this by showing how to visit multiple metric constraints at once and perform a large number of projections simultaneously without conflicts or locking variables.

\subsection{Performing Two Simultaneous Projections}
To develop intuition for our approach, we consider two sets of triplets $t_1 = (a,b,c)$ and $t_2 = (i,j,k)$, where the indices within each triplet are distinct, but some indices may be the same across both triplets. Each of these triplets is associated with three metric constraints, thus three projection steps that must be performed during one pass through the constraint set using Dykstra's method.

Performing projections associate with triplet $t_1$ involves variables $\{ x_{ab}, x_{bc}, x_{ac} \}$. Similarly, triplet $t_2$ is associated with variables $\{ x_{ij}, x_{jk}, x_{ik} \}$. Note that if these triplets share two indices (e.g. $a = i$ and $b = j$), then we cannot perform projections at both constrains in parallel without conflict, since one variable (e.g. $x_{ab} = x_{ij}$) would be updated by both projections. However, if $t_1$ and $t_2$ share at most one index in common, then $\{x_{ab}, x_{bc}, x_{ac}, x_{ij}, x_{jk}, x_{ik} \}$ are all distinct and we can perform projection steps in Dykstra's iteration at $t_1$ and $t_2$ at the same time. Our goal is to use this observation to develop a parallel execution schedule that will allow us to visit a large number of metric constraints at once and perform simultaneous projection steps without conflicts. Because this amounts simply to a re-ordering of constraints in a way that is more easily parallelizable, this will not affect the convergence guarantees of Dykstra's method.

\subsection{New Ordering for Visiting Triplets}
We abstract the process of visiting metric constraints to the process of enumerating triplets of the form $(i,j,k)$ where $1 \leq i < j < k \leq n$. Let $T$ denote this set of ordered triplets. Each fixed ordered triplet will be associated with three different metric constraints, and hence three different projection steps, that we assume will be handled by the same processor in a parallelized projection method.

Based on our intuitive observation in the previous section, we wish to group the triplets in $T$ into subsets $S_1,S_2, \hdots , S_\ell$ in such a way that $S_u \cap S_v = \emptyset$, $\bigcup_{u=1}^\ell S_u = T$, and such that any two triplets in different sets will share at most one index in common. If we can accomplish this, then we can assign each set $S_u$ to a different thread or processor. The work done at each processor (i.e. each set of triplets) will be then completely independent of work performed at other sets by different processors.

To accomplish this we define sets of triplets in which the smallest and largest indices are fixed values $i,k$ such that $k \geq i + 2$. We specifically define
\[ S_{i,k} = \{ (i,j,k) \in T: k \geq i + 2, 1 \leq  i < j < k \leq n  \}\]
which includes all triplets with $i$ as the smallest index and $k$ as the largest index. In Figure~\ref{fig:blocks1} we show a grid of $(i,k)$ pairs associated with $S_{i,k}$ sets. Observe that drawing lines along downward-sloping diagonals of this grid highlights a large number of sets that can be processed simultaneously, i.e any two triplets taken from different sets along the diagonal will share at most one common index. Note that for a fixed $x,z$ satisfying $z \geq x + 2$, the diagonals in Figure~\ref{fig:blocks1} are made up of sets of the form $S_{x+c,z-c}$ for $c = 0, 1, 2, \hdots , \left \lfloor\frac{z - x- 2}{2} \right\rfloor$. The upper bound $c \leq \frac{z - x - 2}{2}$ is chosen to guarantee that $2 + (x + c) \leq (z-c)$, implying that $S_{x+c,z-c}$ contains at least one ordered triplet from $T$.  Based on this observation, in Figure~\ref{for} we show how to loop through all triplets in $T$ in such a way that the inner loop iterates through sets $S_{x+c,z-c}$ that can be processed simultaneously. The code in Figure~\ref{for} contains two double loops for visiting $S_{i,k}$ sets. The first double loop handles the main diagonal of sets in Figure~\ref{fig:blocks1} and everything above below it, and the second double loop iterates through the sets above the main diagonal. Equivalently, for the first double loop we set $x = 1$ and $z \leq n$, and then in the outer loop decrement $z$ by one at each step. The second double loop fixes $z = n$ and iterates through all possibilities $x \in [2, n - 2]$ in the outer loop.

\begin{figure}[h]
	\begin{lstlisting}
Input: integer $n$
Output: triplets $(i,j,k)$ s.t. $1 \leq i < j < k \leq n$
$x = 1$ // First double loop fixes $x$
for $z = n:-1:3$
  $g = floor( (z-x-2)/2)$
  for $c = 0:1:g$	
    $i = x+c$
    $k = z - c$
    List triplets in $S_{i,k}$
    // Different sets of triplets share at most one index
  end
end
$z = n$ // Second double loop fixes $z$
for $x = 2:1:(n-2)$
  $g = floor( (z-x-2)/2)$
  for $c = 0:1:g$
    $i = x+c$
    $k = z - c$
   List triplets in $S_{i,k}$
   // Different sets of triplets share at most one index
  end
end
	\end{lstlisting}
	\vspace{-1em}
	\caption{Loops for listing all triplets in $T$. The inner loops can be perfectly parallelized when we are performing projections at metric constraints of the form $x_{ij} \leq x_{ik} + x_{jk}$, since any two triplets from different $S_{i,k}$ sets will share at most one triplet index in common.}
	\label{for} 
\end{figure}

\begin{figure}
	\centering
	\includegraphics[width=.75\linewidth]{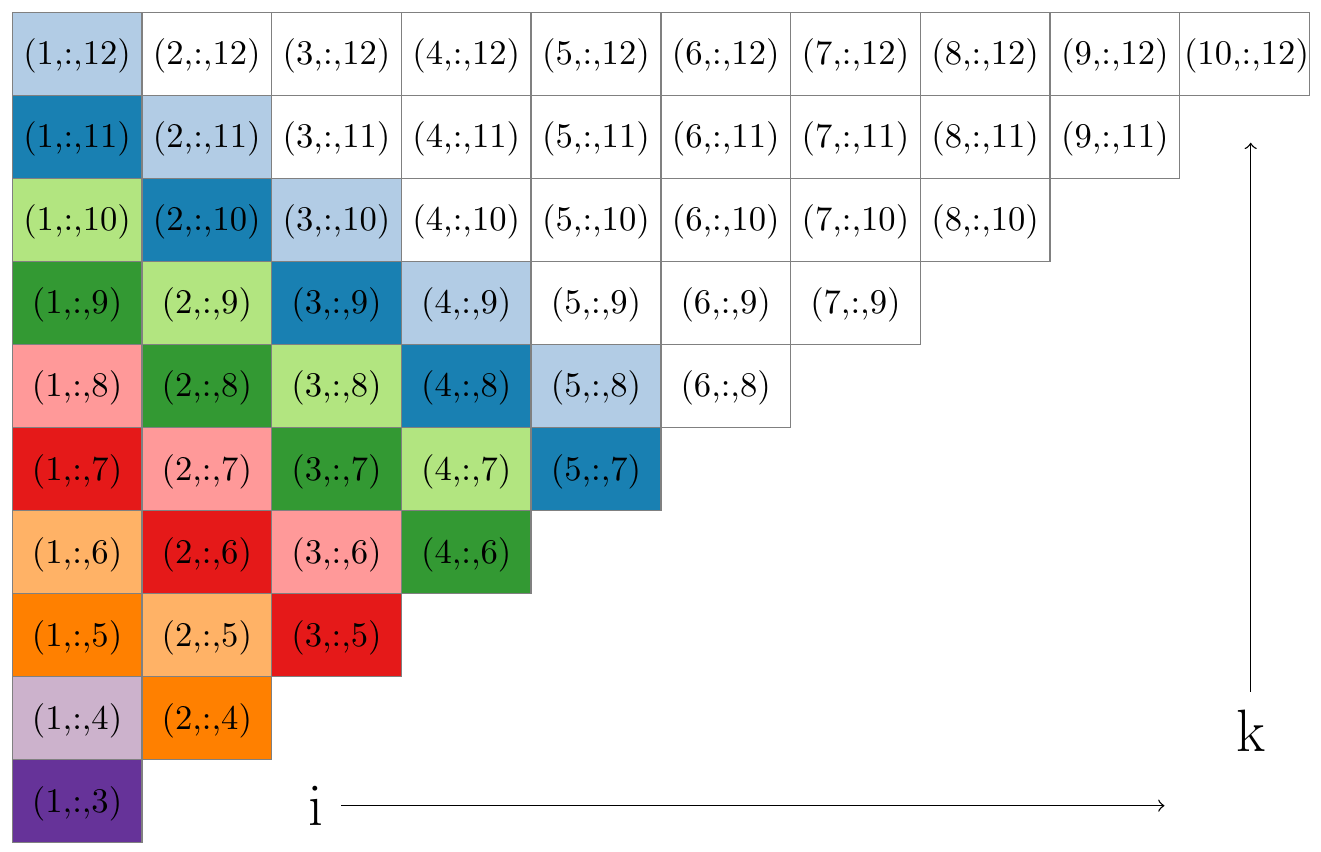}
	\caption{We illustrate how to parallelize visiting metric constraints when $n = 12$. Rectangle $(i,:,k)$ in the grid represents the set of all ordered triplets $S_{i,k}$ for which $i$ is the first index and $k$ is the last index.	When performing projections at metric constraints, sets of the same color can be processed simultaneously without conflict. Each color corresponds to all the $S_{i,k}$ sets that are listed by the inner loop of the first double loop in Figure~\ref{for}, for a fixed $z$. The second double loop in Figure~\ref{for} corresponds to the upper triangular portion of the grid (in white).}
	\label{fig:blocks1}
\end{figure}

\begin{figure}
	\centering
	\includegraphics[width=.75\linewidth]{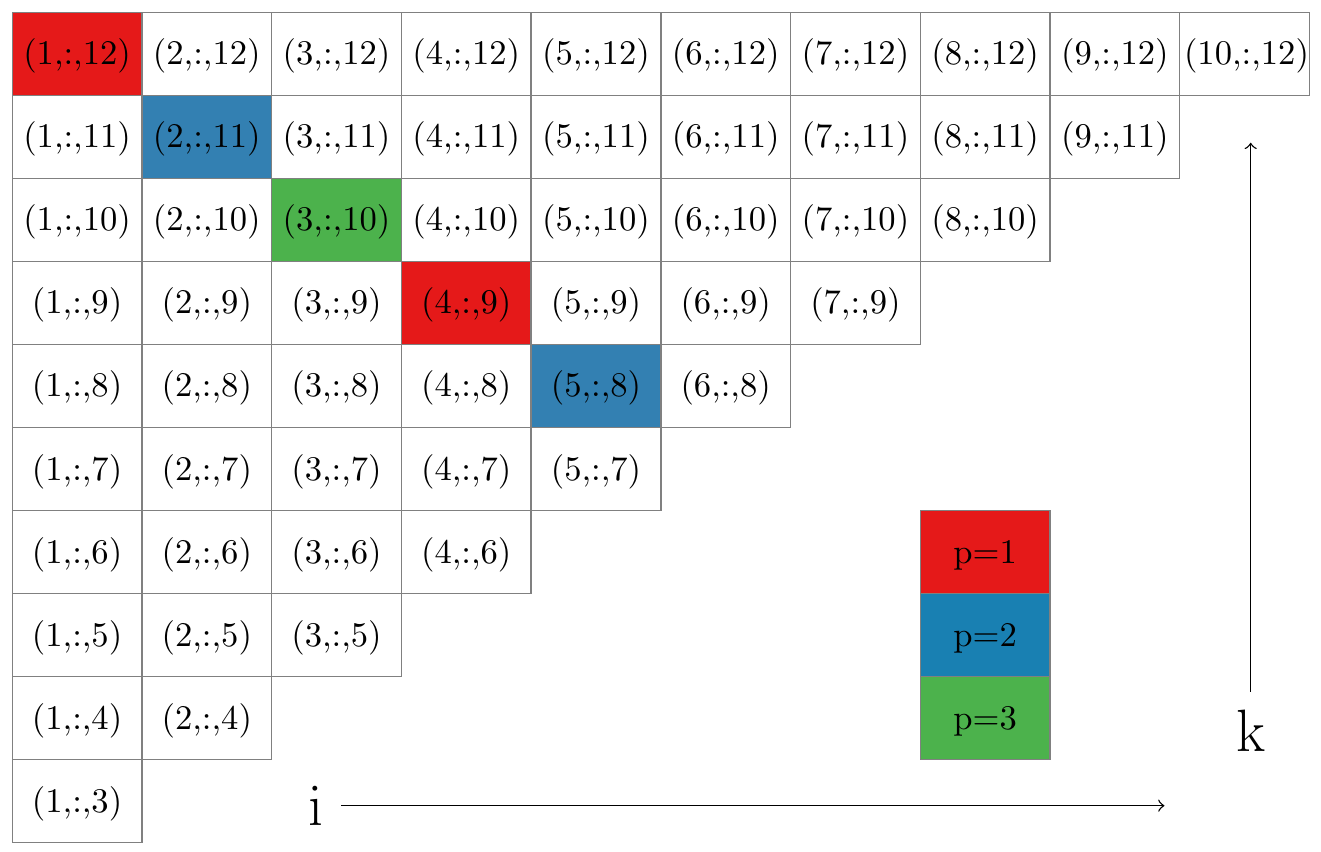}
	\caption{In theory, all sets of triplets along the same diagonal could be handled simultaneously if they were assigned to different processor. In practice, the number processors $p$ is much smaller than $n$. We balance the load among processors by assigning the $r$th set on a diagonal to processor $r\mod p$ (where processor 0 and processor $p$ are the same). Here we illustrate the assignment of triplet sets to processors along the main diagonal when $n = 12$ and $p = 3$.}
	\label{fig:blocks2}
\end{figure} 

\subsection{Load Balancing and Tiled Triplet Assignment}
One issue we must address with our listing of triplets in Figure~\ref{for} is the load balance. There is variability in both the number of triplet sets in the parallelized inner loop (i.e. the number of entries along a given diagonal in Figure~\ref{fig:blocks1}), as well as the size of each triplet set within the same inner loop (i.e. different entries in the same diagonal of the grid in Figure~\ref{fig:blocks1}). For example, when $x = 1$ and for a fixed $z$, there are $(z-3)/2 +1$ sets, and these sets have variable size $z-2(c+1)$ for different values of $c$ ranging from $0$ to $(z-3)/2)$. We begin by noting that the vast majority of the triplets are visited for values of $z = O(n)$. Secondly, we assume that the number of threads or processors $p$ we use when iterating over sets is significantly smaller than the problems size $n$, and we can assign sets of triplets to processors in a way that will not be too imbalanced. For a diagonal defined by fixed $x,z$ values, there are $\left \lfloor\frac{z - x- 2}{2} \right\rfloor$ sets of triplets. If we assigned the first group of $n/p$ triplet sets to the first processor, and in general assigned the $r$th group of $n/p$ triplet sets to the $r$th processor, this would indeed lead to a significant imbalance. However, in practice, we balance the load much more effectively by assigning the $r$th set $S_{i,k}$ to processor $r \mod p$. In this way each processor is responsible for different triplet sets with a range of different sizes, for an overall load that is roughly balanced. We illustrate this load balanced assignment in Figure~\ref{fig:blocks2}.

\begin{figure}
	\centering
	\includegraphics[width=.9\linewidth]{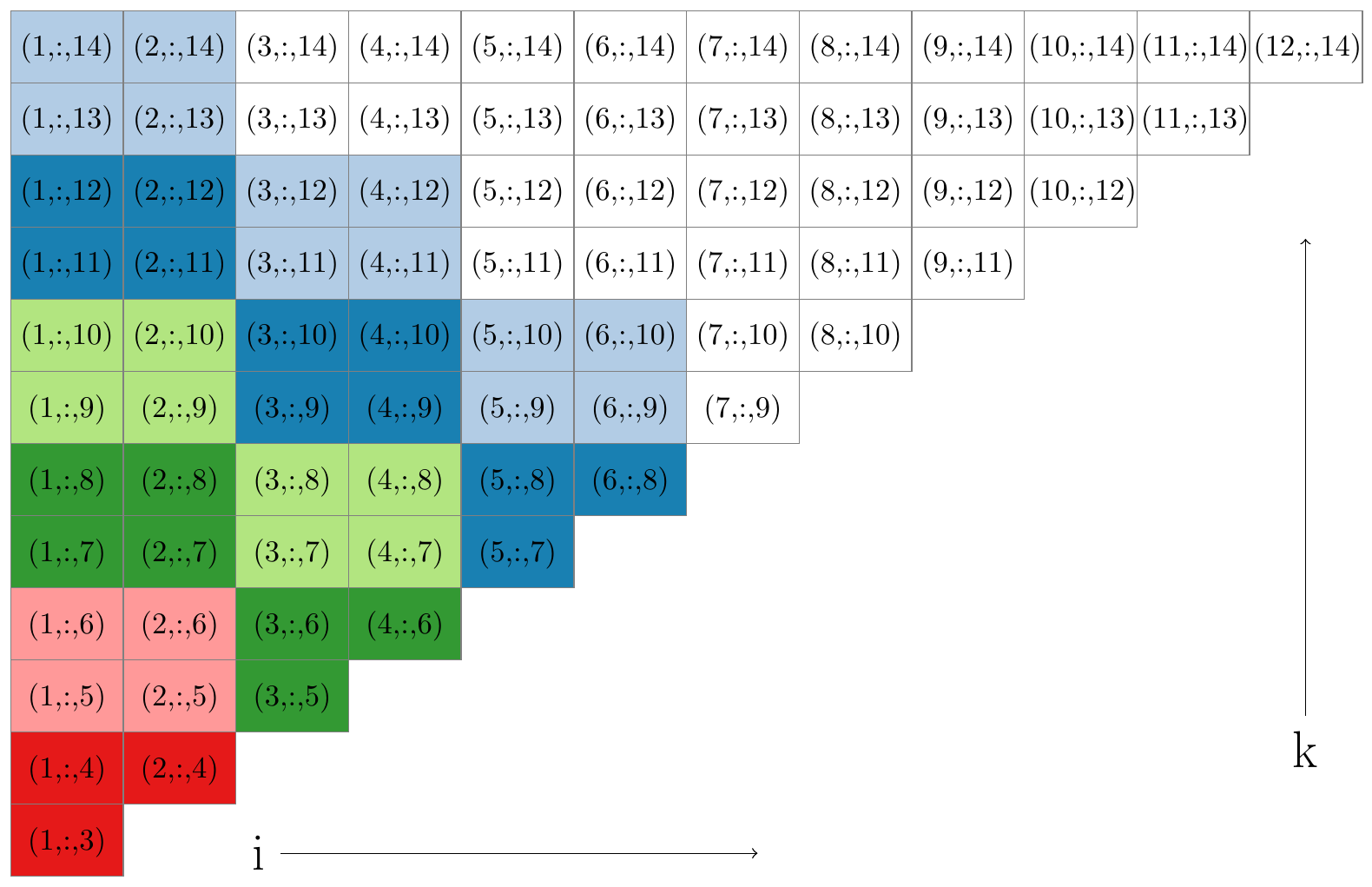}
	\caption{We illustrate the tiling approach for visiting triplet sets when $n = 14$ and tile size $b = 2$. Triplets in different tiles along the same diagonal can be visited simultaneously without conflict when assigned to different processors. Organizing sets $S_{i,k}$ into $b \times b$ tiles and carefully iterating through middles indices $j$ allows for better cache efficiency when accessing variables in the matrix $\mX$.}
	\label{fig:tiled}
\end{figure}

We also improve our parallel execution schedule by implementing a tiled approach to triplet set assignment for better cache efficiency when accessing distance variable in the matrix $\mX = (x_{ij})$. This is inspired by previous work on tiled matrix multiplication, though it differs slightly in order to apply to enumerating triplets specifically for metric constrained projection methods. In short, this tiled strategy corresponds to substituting the diagonal pattern in Figure~\ref{fig:blocks1} with the block diagonal pattern shown in Figure~\ref{fig:tiled}. In more detail, for a tile size $b$, each tile is defined by a fixed $(x,z)$ pair, and is made up of all $S_{i,k}$ sets where $i \in \{x, x+1 \hdots , x+b-1 \}$ and $k \in \{z, z-1, \hdots , z-b+1 \}$. Much like in the untiled case, we note that different tiles of the same color in Figure~\ref{fig:tiled} can be visited by different processors at the same time without conflict. That is, different processors will access completely independent parts of the matrix $\mX$. When assigning $p$ processors to tiles along a block diagonal, we assign the $r$th tile to processor $r \mod p$, generalizing the strategy outlined in Figure~\ref{fig:blocks2} for the untiled case.

\begin{figure}[tb]
	\centering
	\subfigure{%
		\includegraphics[width=1.5in]{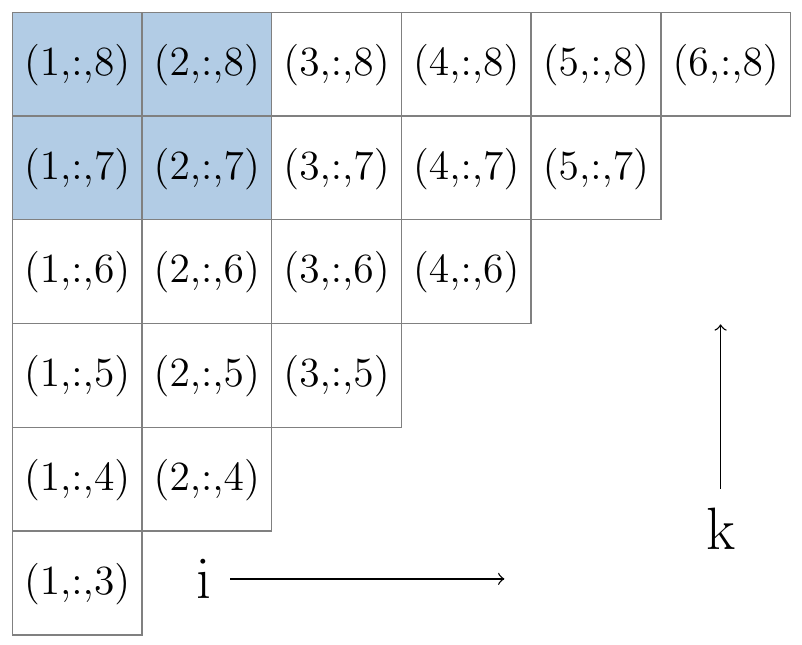}
		\label{fig:n8}} $\qquad$
	\subfigure{%
		\includegraphics[width=1.5in]{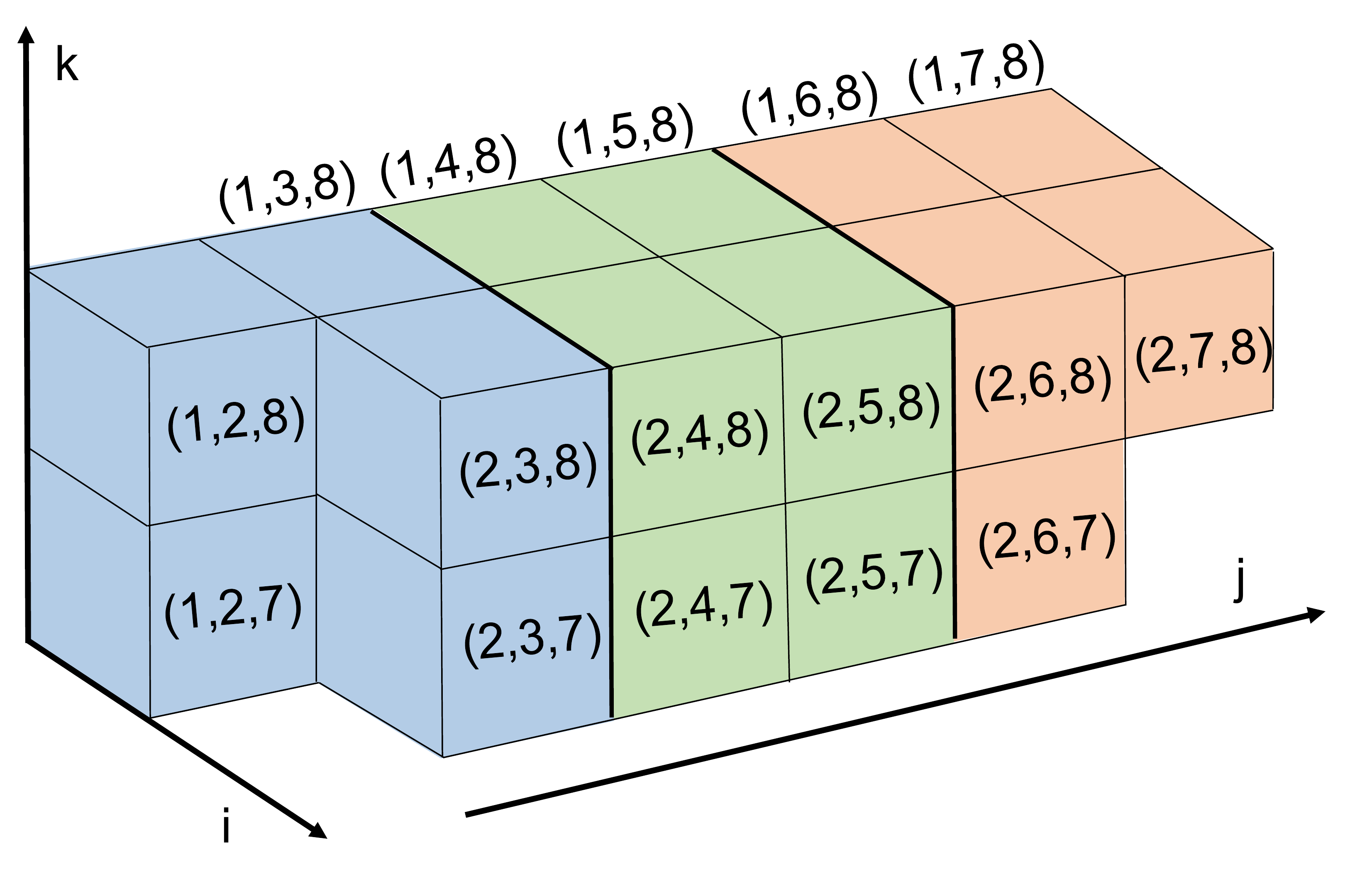}
		\label{fig:cube}}
	\caption{Each tile in the $(i,k)$ grid is assigned to different processor. The blue section in the left image corresponds to a single tile when $b = 2$ and $n = 8$. When a processor is assigned such a tile, it also separates options for the middle index $j$ into groups of $b$ choices. The processor then iterates through triplets $(i,j,k)$ in cubes of size $b \times b \times b$ in a way that maximizes column locality and cache efficiency when accessing entries of the form $x_{ij}, x_{ik}, x_{jk}$ from the distance matrix $\mX$. The right figure highlights this process specifically for the tile in the figure on the left. The processor handles each colored section in turn. Some of the cubes are incomplete, since triplets such as $(2,2,8)$, $(2,2,7)$, and $(2,7,7)$ do not satisfy $i < j < k$.}
	\label{fig:tiles2}
\end{figure}

Each tile, which is defined by a fixed pair $(x,z)$ and a tile size $b$, is associated with $b$ choices for the smallest index $i$ and $b$ choices for the largest index $k$. The processor assigned to this tile must then iterate through all valid middle indices $j$ and perform projections corresponding to triplets of the form $(i,j,k)$. One approach for doing this would be to consider each $(i,k)$ pair in turn and iterate through all values of $j$ from $j= i+1$ to $j = k-1$. However, for better cache efficiency, we instead split the full range of possible $j$ values from $x+1$ to $z-1$ into subintervals that are also of length $b$. This gives us a sequence of $b \times b \times b$ cubes of $(i,j,k)$ values, each associated with entries $x_{ij}$, $x_{ik}$, and $x_{jk}$ from $\mX$. Within each of these cubes, we iterate through triplets in a way that maximizes column locality (assuming $\mX$ is stored in column major format), before moving on to the next cube. We give a simple illustration of this in Figure~\ref{fig:tiles2}.
 Depending on the values of $x$,$z$, and $b$, we will of course not be able to organize all triplets into perfect $b \times b \times b$ cubes of $(i,j,k)$ triplets satisfying $i < j < k$. In practice however for large values of $n$ and $b \ll n$, this approach will still provide a balanced and cache efficient way to access variables in $\mX$.

\subsection{Storing Dual Variables for Parallel Computations}
In addition to updating primal variables $(x_{ij})$, Dykstra's method requires we keep track of dual variables as a part of correction step (line 7 in Algorithm 1) that is necessary to guarantee convergence to the optimal solution. Specifically, for each metric constraint associated with a triplet $(i,j,k)$, there is a corresponding dual variable $y_{ijk}$ that is updated during each visit to a constraint. This variable is only nonzero if in the previous pass through the constraints, there was a non-trivial projection step (i.e. the entries $x_{ij}$, $x_{ik}$, $x_{jk}$ changed). For serial projection methods for metric optimization, the metric constraints are visited in the same order in every pass through the constraint set~\cite{veldt2018projection}. This makes it possible to effectively query dual variables from an array that stores tuples of the form $(t_{ijk},y_{ijk})$ where $t_{ijk}$ is a unique index associated with a metric constraint and $y_{ijk}$ is the dual variable. For memory-efficiency, these tuples are only stored for nonzero dual variables: $y_{ijk} > 0$. Because the serial version visits constraints in the same order each round, the array is always traversed in the same order. At each step, the method can access each necessary dual variable in $O(1)$ time by maintaining a pointer in the array to the next known triplet $(i,j,k)$ associated with a nonzero dual variable.
 In this way the method can access the necessary dual variables in $O(1)$ time. 

Using our new parallel execution schedule, the triplets are no longer visited in a deterministic fashion, so a new approach is necessary. Fortunately, our approach is designed in such a way that each triplet $(i,j,k)$ is always visited by the same processor during each different pass through the constraints. Furthermore, even though globally the triplets are not visited in a deterministic fashion, each individual processor visits its assigned triplets in the same deterministic order at every iteration. Therefore, we can maintain dual variables efficiently by assigning an array to each processor, allowing the processor to keep track of the next triplet it will visit that will require a non-trivial correction step. Thus the main difference between the serial and parallel versions is simply that the latter requires we maintain an array for each processor rather than a single array for storing all dual variables. Accessing dual variables is therefore still performed in $O(1)$ time at each projection step and the theoretical memory complexity is the same.

\section{Experiments}


We demonstrate the power of our new parallel approach to metric-constrained optimization by using it to solve the linear programming relaxation of correlation clustering on several large instances. We find that using even a modest number of cores consistently leads to a speed up of roughly a factor 5, and up to a speedup over a factor 10 on the largest problem, which involves nearly 3 trillion constraints.

\subsection{Implementation Details}
We implement a solver for the metric-constrained LP relaxation of correlation clustering by incorporating our new parallel execution schedule into our previous serial framework~\cite{veldt2018projection}. We use the Julia programming language, using its support for threaded computations to parallelize the inner loops of our new approach to iterating through index triplets. Our code is available publicly online at~\url{https://github.com/camruggles/ParallelDykstras}. In our experiments we compare against our previous serial projection methods, available at at~\url{https://github.com/nveldt/MetricOptimization}. 

\subsection{Problem Construction and Datasets}
To test our parallel solver we construct several large instances of correlation clustering from undirected graphs following the approach of Wang et al.~\cite{wang2013scalable}, and including a slight modification applied in previous work~\cite{veldt2018projection}. In short, given a graph $G = (V,E)$, we compute a signed and weighted edge between each pair of nodes $(i,j)$ by computing the Jaccard index between the nodes (which is always nonnegative) and applying a non-linear function to obtain a signed value that either represents similarity or dissimilarity between the nodes. We then offset these scores by $\pm \varepsilon$ for a small $\varepsilon > 0$. This last step ensures the result will be an instance of correlation clustering in which each pair of nodes possesses a nonzero weight and a sign. Partitioning the original graph $G$ using the correlation clustering objective can be used as a way to perform community detection on $G$. For our purposes, this construction leads to a dense instance of correlation clustering that serves as a good benchmark for solving the LP relaxation of correlation clustering on a large scale.

We apply this procedure to five undirected and unsigned graphs: the graph \emph{power} from the Newman group of matrices in the SuiteSparse Matrix Collection~\cite{power,suitesparse}, and four collaboration networks available from the SNAP repository~\cite{snapnets,leskovec2007}: \emph{ca-GrQc}, \emph{ca-HepTh}, \emph{ca-HepPh}, and \emph{ca-AstroPh}. We take the largest connected component of each graph before converting it into an instance of correlation clustering. The LP relaxation of the correlation clustering instance corresponding to the largest graph (\emph{ca-AstroPh}) has over 160 million variables and 2.9 trillion constraints.

\subsection{Machine Specifications and Computing Environment}
Our experiments were almost exclusively performed on a computer with 4 16-core Intel Xeon E7-8867 v3 processors. For one experiment on the largest graph, in which we wanted to run a large number of cores, we used a machine with 8 24-core 2.7 GHz Intel Xeon Platinum 8168 processors.
For our experiments we did not utilize
exclusive access to the computers.
Thus, the reported runtimes vary
depending on whether there were other users simultaneously
using the machine at the same time as our experiments.
This emulates the natural and realistic performance
that may be expected in settings such as Amazon EC2,
with multiple shared VMs on a single machine.

%

\subsection{The Effect of Reordering Constraints}
Dykstra's method is guaranteed to converge regardless of the order in which the constraints are visited. However, we found that in practice the number of iterations required to solve a problem to within a fixed tolerance for constraint satisfaction and duality gap did vary depending on the constraint ordering. In some cases, the standard serial ordering led to a smaller overall iteration count, though in many other cases the iteration count was lower for our new approach for visiting triplets. Given the variability between problem instances, in our experiments we focus simply on the time it takes to complete a fixed number of iterations of Dykstra's method. In this way, we are always comparing the time it takes to visit and perform a step of Dykstra's method at each individual constraint exactly $C$ times for some fixed integer $C$.  

\subsection{Results}
In Table~\ref{runtimes}, we report results for running our parallel code on all five graphs using 8, 16, and 32 cores. The runtime for 1 core comes from applying the previous serial version of the algorithm~\cite{veldt2018projection}.
For the largest graph we additionally run our new algorithm using 64 cores, which is the only experiment for which we used the machine with 8 24-core processors. For each graph we report the time it took in seconds to run Dykstra's method for 20 iterations, using a tile size of $b = 40$. Running our method with 8 cores is consistently 4-5 times faster than the serial implementation. We continue to see performance gains as we increase the number of cores used, leading to a speedup of over a factor ten on our largest graph.
	\begin{table}[h]
		\caption{Results for parallel Dykstra's method in solving the metric-constrained LP relaxation of correlation clustering.
			}			
		\label{runtimes}
		\centering
		\begin{tabular}{ll  c c c}
			\toprule
			Graph & \# constraints & \# Cores & Times (s)  & Speedup \\
			 \midrule
			ca-GrQc &$3.6 \times 10^{10}$ & 1& 2632&1 \\
		 $n = 4158$	& & 8 & 562 & 4.68 \\
			& & 16 & 429 & 6.14 \\
			& & 32& 358 & 7.35\\
			\midrule
			Power  & $6.0 \times 10^{10}$ & 1&4521 & 1\\
			$n = 4941$ & & 8& 890& 5.08 \\
			& &16 & 696& 6.50 \\
			& &32 & 576 & 7.85\\
			\midrule
			 ca-HepTh  &$3.2 \times 10^{11}$ & 1 & 19826& 1\\
			$n =8638$ & & 8& 4682 & 4.23 \\
			& & 16 &3252& 6.10  \\
			& & 32& 2603 & 7.62 \\
			\midrule						 
			ca-HepPh  & $7.0 \times 10^{11}$& 1 & 47309 & 1\\
			$n = 11204$ & & 8& 10313& 4.59\\
			& & 16 & 7066&  6.70\\
			& & 32& 5889& 8.03 \\
			\midrule		
			ca-AstroPh & $2.9 \times 10^{12}$ & 1 & 187045  & 1\\
			$n = 17903$ & &8 & 40146 & 4.66 \\
			&& 16 & 35397& 5.28\\
			& & 32& 24374 & 7.67 \\
			& & 64 & 16325 & 11.46 \\	
		\bottomrule
		\end{tabular}
	\end{table}
	
\begin{figure}
			\centering
			\includegraphics[width=.75\linewidth]{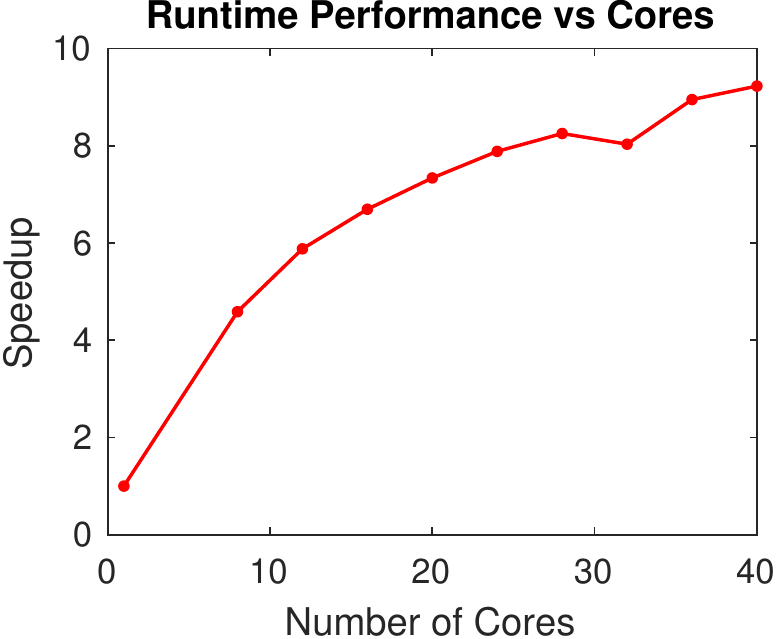}
			\caption{We display results for varying the number of cores on \emph{ca-HepPh} for a fixed tile size of 40. Results are displayed for 1 core, and then for 8 to 40 cores in increments of 4.}
			\label{fig:threadvary}
		\end{figure}

\begin{figure}
		\centering
		\includegraphics[width=.75\linewidth]{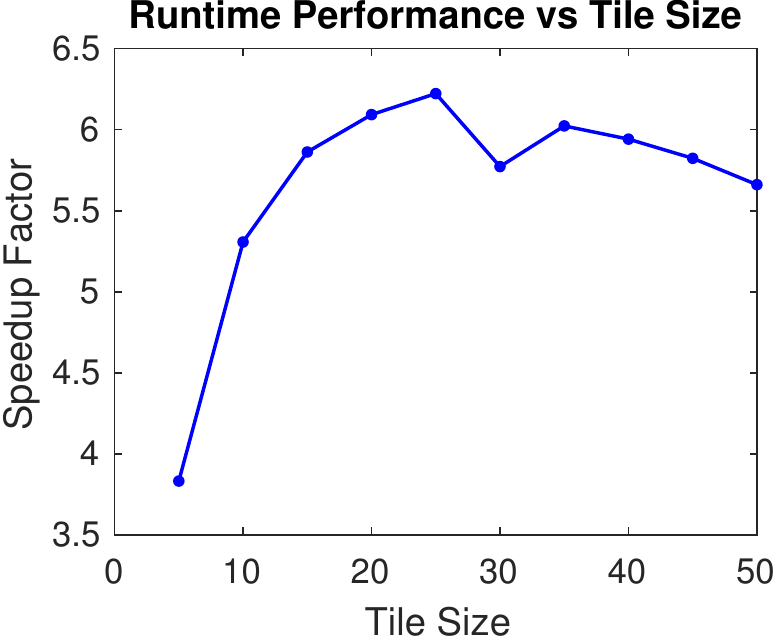}
		\caption{When we vary the tile size, performance climbs to a peak and then slowly decreases if we increase the tile size too much. Results are shown here for graph \emph{ca-GrQc} when tile size ranges from 5 to 50 in increments of 5. The number of threads is fixed at 16.}
		\label{fig:tilevary}
	\end{figure}

	In Figure~\ref{fig:threadvary} we display results specifically on \emph{ca-HepPh} using a wider range of core counts. We see the performance of our method increase sharply at first and slowly level off as we increase the number of cores.
		Finally, we observe what happens as we vary the tile size and keep the number of cores fixed. Figure~\ref{fig:tilevary} illustrates the algorithm's performance on \emph{ca-GrQc} as we vary tile size from 5 to 50 and keep the number of cores fixed at 16. The curve in the figure shows the speedup over the serial implementation, which is above a factor 5 for all tile sizes except 5. The performance peaks just above a factor 6 speedup for a tile size of 25, and slowly begins to decrease after this point. 

\section{Related Work}

Our work is related to a number of different areas in machine learning, optimization, graph theory, and matrix computations.

\paragraph{Metric-Optimization and Projection Methods}
 The parallel algorithms we have develop build directly on previous serial techniques for metric constrained optimization. These optimization problems arise in algorithm design for graph clustering problems~\cite{charikar2005clustering,leighton1999multicommodity,Agarwal2008metricmod,veldt2017lamcc}, image segmentation~\cite{glasner2011contour,Vitaladevuni2010coclustering}, sensor location~\cite{gentile2005sensor,gentile2007distributed}, metric nearness~\cite{brickell2008metricnearness}, and metric learning~\cite{biswas2014semi,batra2008semi}. Sra et al.~\cite{sra2005MNnips} were the first to apply projection methods for metric-constrained optimization, by using Dykstra's method~\cite{dykstra1983algorithm} to solve different variants of metric nearness. In recent work~\cite{veldt2018projection}, we developed improved techniques for applying this method more broadly to linear programming relaxations of graph clustering objectives. 

\paragraph{Graph Coloring}
The parallel execution schedule we have develop in this paper is related to a number of different graph coloring problems. Consider a graph in which every node corresponds to a triplet $(i,j,k)$, and edges connect nodes (i.e. triplets) if they share two indices in common. Coloring the nodes in this graph in such a way that no adjacent nodes share the same color is equivalent to partitioning all triplets into disjoint sets such that triplets within the same set can be processed simultaneously by our projection methods. Another approach would be to instead assign each pair $ij$ (i.e. each entry in the distance matrix $\mX = (x_{ij})$) to a node in a hypergraph, and for every triplet of indices $(i,j,k)$ define a hyperedge of the form $(ij,jk,ik)$. Then the problem of finding sets of triplets to process simultaneous is equivalent to edge coloring in 3-uniform hypergraphs ~\cite{Obszarski2017edgecoloring}. In general, graph coloring arises frequently as a way to determine potential areas for concurrency when completing a given task in parallel. We refer to several helpful resources on coloring algorithms for parallel and multithreaded computations~\cite{10.1007/978-3-642-23397-5_24,CATALYUREK2012576,Gebremedhin:2013:CSG:2513109.2513110}.

\paragraph{Block Matrix Multiplication}
Our tiled approach to triplet enumeration is inspired by techniques for block matrix multiplication, which also involves doubly indexed blocks of data and computational steps corresponding to a triplet of indices. Specifically, multiplying the $ij$ block of a matrix $\mA$ with the $jk$ block of another matrix $\mB$ is a step in block matrix-matrix multiplication ($\mA\mB = \mC$), that can be indexed by a triplet $(i,j,k)$. Our tiled triplet enumeration procedure is related to research on communication bounds for dense matrix multiplication. The pioneering work of Hong and Kung~\cite{jia1981complexity} proved a lower bound on the communication necessary to move data between slow and fast memory in matrix multiplication. Irony, Toledo, and Tiskin~\cite{irony2004communication} later extended this result to distributed parallel computations. The state of the art numerical linear algebra software package LAPACK~\cite{laug} determines block sizes automatically for efficient matrix-matrix computations. For an in-depth overview of communication-avoiding and cache efficient algorithms for numerical linear algebra, we refer to the work of Ballard et al.~\cite{ballard2014communication} and Knight~\cite{knight2015communication} (see in particular Section 5.5).

%
%

\section{Discussion and Future Work}

In our work we have taken a first step in developing parallel algorithms for metric-constrained optimization. These problems are very challenging to solve in practice due to their extremely large constraint set, involving $O(n^3)$ constraints for a dataset of size $n$. Furthermore, parallelizing solvers for these problems possess several very significant challenges, including the P-completeness of linear programming and the downsides of applying existing parallel versions of projection methods. Despite this, we have demonstrated that exploiting the special structure of the constraint matrix can lead to noticeable performance improvements, in particular when applying projection methods such as Dykstra's method. Our work demonstrates that metric-constrained optimization problems are challenging to solve, but also serve as good benchmarks for testing parallel design techniques. In future work we will continue to explore other even more effective ways to visit metric constraints in parallel, as well as other possible ways to exploit the special structure of the constraint matrix in order to solve these problems more effectively in practice.

\bibliographystyle{plain}
\bibliography{triconstraints}  

\begin{thebibliography}{10}

\bibitem{Agarwal2008metricmod}
G.~Agarwal and D.~Kempe.
\newblock Modularity-maximizing graph communities via mathematical programming.
\newblock {\em The European Physical Journal B}, 66(3):409--418, Dec 2008.

\bibitem{AilonCharikarNewman2008}
Nir Ailon, Moses Charikar, and Alantha Newman.
\newblock Aggregating inconsistent information: ranking and clustering.
\newblock {\em Journal of the ACM (JACM)}, 55(5):23, 2008.

\bibitem{laug}
E.~Anderson, Z.~Bai, C.~Bischof, S.~Blackford, J.~Demmel, J.~Dongarra,
  J.~Du~Croz, A.~Greenbaum, S.~Hammarling, A.~McKenney, and D.~Sorensen.
\newblock {\em {LAPACK} Users' Guide}.
\newblock Society for Industrial and Applied Mathematics, Philadelphia, PA,
  third edition, 1999.

\bibitem{ballard2014communication}
Grey Ballard, Erin Carson, James Demmel, Mark Hoemmen, Nicholas Knight, and
  Oded Schwartz.
\newblock Communication lower bounds and optimal algorithms for numerical
  linear algebra.
\newblock {\em Acta Numerica}, 23:1--155, 2014.

\bibitem{Bansal2004correlation}
Nikhil Bansal, Avrim Blum, and Shuchi Chawla.
\newblock Correlation clustering.
\newblock {\em Machine Learning}, 56:89--113, 2004.

\bibitem{batra2008semi}
D.~Batra, R.~Sukthankar, and T.~Chen.
\newblock Semi-supervised clustering via learnt codeword distances.
\newblock In {\em Proceedings of the British Machine Vision Conference}, BMVA
  2008, pages 90.1--90.10. BMVA Press, 2008.
\newblock doi:10.5244/C.22.90.

\bibitem{biswas2014semi}
Arijit Biswas.
\newblock {\em Semi-supervised and Active Image Clustering with Pairwise
  Constraints from Humans}.
\newblock PhD thesis, University of Maryland, College Park, 2014.

\bibitem{brickell2008metricnearness}
Justin Brickell, Inderjit~S. Dhillon, Suvrit Sra, and Joel~A. Tropp.
\newblock The metric nearness problem.
\newblock {\em SIAM Journal on Matrix Analysis and Applications},
  30(1):375--396, 2008.

\bibitem{CATALYUREK2012576}
{\"U}mit~V {\c{C}}ataly{\"u}rek, John Feo, Assefaw~H Gebremedhin, Mahantesh
  Halappanavar, and Alex Pothen.
\newblock Graph coloring algorithms for multi-core and massively multithreaded
  architectures.
\newblock {\em Parallel Computing}, 38(10-11):576--594, 2012.

\bibitem{charikar2005clustering}
Moses Charikar, Venkatesan Guruswami, and Anthony Wirth.
\newblock Clustering with qualitative information.
\newblock {\em Journal of Computer and System Sciences}, 71(3):360 -- 383,
  2005.
\newblock Learning Theory 2003.

\bibitem{chawla2015near}
Shuchi Chawla, Konstantin Makarychev, Tselil Schramm, and Grigory Yaroslavtsev.
\newblock Near optimal {LP} rounding algorithm for correlation clustering on
  complete and complete {$k$}-partite graphs.
\newblock In {\em Proceedings of the Forty-Seventh Annual ACM on Symposium on
  Theory of Computing}, STOC 2015, pages 219--228. ACM, 2015.

\bibitem{suitesparse}
Timothy~A. Davis and Yifan Hu.
\newblock The university of florida sparse matrix collection.
\newblock {\em ACM Trans. Math. Softw.}, 38(1):1:1--1:25, December 2011.

\bibitem{DemaineEmanuelFiatEtAl2006}
Erik~D. Demaine, Dotan Emanuel, Amos Fiat, and Nicole Immorlica.
\newblock Correlation clustering in general weighted graphs.
\newblock {\em Theoretical Computer Science}, 361(2):172 -- 187, 2006.
\newblock Approximation and Online Algorithms.

\bibitem{dhillon2003MNreport}
Inderjit~S Dhillon, Suvrit Sra, and Joel~A Tropp.
\newblock The metric nearness problems with applications.
\newblock Technical report, 2003.

\bibitem{dhillon2004tfa}
Inderjit~S. Dhillon, Suvrit Sra, and Joel~A. Tropp.
\newblock Triangle fixing algorithms for the metric nearness problem.
\newblock In {\em Advances in Neural Information Processing Systems 17}, NIPS
  2004, pages 361--368, Cambridge, MA, USA, 2004. MIT Press.

\bibitem{dykstra1983algorithm}
Richard~L Dykstra.
\newblock An algorithm for restricted least squares regression.
\newblock {\em Journal of the American Statistical Association},
  78(384):837--842, 1983.

\bibitem{escalante2011altproj}
R.~Escalante and M.~Raydan.
\newblock {\em Alternating Projection Methods}.
\newblock Society for Industrial and Applied Mathematics, Philadelphia, PA,
  2011.

\bibitem{Gebremedhin:2013:CSG:2513109.2513110}
Assefaw~H. Gebremedhin, Duc Nguyen, Md. Mostofa~Ali Patwary, and Alex Pothen.
\newblock Colpack: Software for graph coloring and related problems in
  scientific computing.
\newblock {\em ACM Trans. Math. Softw.}, 40(1):1:1--1:31, October 2013.

\bibitem{gentile2005sensor}
Camillo Gentile.
\newblock Sensor location through linear programming with triangle inequality
  constraints.
\newblock In {\em IEEE International Conference on Communications}, volume~5,
  pages 3192--3196. IEEE, 2005.

\bibitem{gentile2007distributed}
Camillo Gentile.
\newblock Distributed sensor location through linear programming with triangle
  inequality constraints.
\newblock {\em IEEE transactions on wireless communications}, 6(7), 2007.

\bibitem{glasner2011contour}
D.~Glasner, S.~N. Vitaladevuni, and R.~Basri.
\newblock Contour-based joint clustering of multiple segmentations.
\newblock In {\em Proceedings of the 2011 IEEE Conference on Computer Vision
  and Pattern Recognition}, CVPR 2011, pages 2385--2392, Washington, DC, USA,
  2011. IEEE Computer Society.

\bibitem{han1988successive}
Shih-Ping Han.
\newblock A successive projection method.
\newblock {\em Mathematical Programming}, 40(1-3):1--14, 1988.

\bibitem{hildreth1957quadratic}
Clifford Hildreth.
\newblock A quadratic programming procedure.
\newblock {\em Naval Research Logistics (NRL)}, 4(1):79--85, 1957.

\bibitem{jia1981complexity}
Jia-Wei Hong and Hsiang-Tsung Kung.
\newblock I/o complexity: The red-blue pebble game.
\newblock In {\em Proceedings of the thirteenth annual ACM symposium on Theory
  of computing}, STOC '81, pages 326--333. ACM, 1981.

\bibitem{irony2004communication}
Dror Irony, Sivan Toledo, and Alexander Tiskin.
\newblock Communication lower bounds for distributed-memory matrix
  multiplication.
\newblock {\em Journal of Parallel and Distributed Computing},
  64(9):1017--1026, 2004.

\bibitem{iusem1991convergence}
Alfredo~N Iusem and Alvaro~R De~Pierro.
\newblock On the convergence of {H}an's method for convex programming with
  quadratic objective.
\newblock {\em Mathematical Programming}, 52(1-3):265--284, 1991.

\bibitem{knight2015communication}
Nicholas~Sullender Knight.
\newblock {\em Communication-Optimal Loop Nests}.
\newblock PhD thesis, UC Berkeley, 2015.

\bibitem{leighton1999multicommodity}
Tom Leighton and Satish Rao.
\newblock Multicommodity max-flow min-cut theorems and their use in designing
  approximation algorithms.
\newblock {\em Journal of the ACM (JACM)}, 46(6):787--832, November 1999.

\bibitem{leskovec2007}
Jure Leskovec, Jon Kleinberg, and Christos Faloutsos.
\newblock Graph evolution: Densification and shrinking diameters.
\newblock {\em ACM Trans. Knowl. Discov. Data}, 1(1), March 2007.

\bibitem{snapnets}
Jure Leskovec and Andrej Krevl.
\newblock {SNAP Datasets}: {Stanford} large network dataset collection.
\newblock \url{http://snap.stanford.edu/data}, June 2014.

\bibitem{mangasarian1984normal}
O.~L. Mangasarian.
\newblock Normal solutions of linear programs.
\newblock {\em Mathematical Programming at Oberwolfach II}, pages 206--216,
  1984.

\bibitem{Obszarski2017edgecoloring}
Pawe\l{l}l Obszarski and Andrzej Jastrz\c{e}bski.
\newblock Edge-coloring of 3-uniform hypergraphs.
\newblock {\em Discrete Applied Mathematics}, 217:48 -- 52, 2017.
\newblock Combinatorial Optimization: Theory, Computation, and Applications.

\bibitem{10.1007/978-3-642-23397-5_24}
Md. Mostofa~Ali Patwary, Assefaw~H. Gebremedhin, and Alex Pothen.
\newblock New multithreaded ordering and coloring algorithms for multicore
  architectures.
\newblock In Emmanuel Jeannot, Raymond Namyst, and Jean Roman, editors, {\em
  Euro-Par 2011 Parallel Processing}, pages 250--262, Berlin, Heidelberg, 2011.
  Springer Berlin Heidelberg.

\bibitem{puleo2015cc}
{Gregory J.} Puleo and Olgica Milenkovic.
\newblock Correlation clustering with constrained cluster sizes and extended
  weights bounds.
\newblock {\em SIAM Journal on Optimization}, 25(3):1857--1872, 2015.

\bibitem{puleo2016cc}
Gregory~J. Puleo and Olgica Milenkovic.
\newblock Correlation clustering and biclustering with locally bounded errors.
\newblock In {\em Proceedings of the 33rd International Conference on
  International Conference on Machine Learning}, ICML 2016, pages 869--877.
  JMLR.org, 2016.

\bibitem{sra2005MNnips}
Suvrit Sra, Joel Tropp, and Inderjit~S. Dhillon.
\newblock Triangle fixing algorithms for the metric nearness problem.
\newblock In L.~K. Saul, Y.~Weiss, and L.~Bottou, editors, {\em Advances in
  Neural Information Processing Systems}, pages 361--368. MIT Press, 2005.

\bibitem{veldt2018projection}
Nate Veldt, David Gleich, Anthony Wirth, and James Saunderson.
\newblock A projection method for metric-constrained optimization.
\newblock {\em arXiv preprint arXiv:1806.01678}, 2018.

\bibitem{veldt2017lamcc}
Nate Veldt, David~F. Gleich, and Anthony Wirth.
\newblock A correlation clustering framework for community detection.
\newblock In {\em Proceedings of the 2018 World Wide Web Conference}, WWW 2018,
  pages 439--448, 2018.

\bibitem{Vitaladevuni2010coclustering}
S.~N. Vitaladevuni and R.~Basri.
\newblock Co-clustering of image segments using convex optimization applied to
  em neuronal reconstruction.
\newblock In {\em 2010 IEEE Computer Society Conference on Computer Vision and
  Pattern Recognition}, CVPR 2010, pages 2203--2210, June 2010.

\bibitem{wang2013scalable}
Yubo Wang, Linli Xu, Yucheng Chen, and Hao Wang.
\newblock A scalable approach for general correlation clustering.
\newblock In {\em International Conference on Advanced Data Mining and
  Applications}, ADMA 2013, pages 13--24. Springer, 2013.

\bibitem{power}
Duncan~J. Watts and Steven~H. Strogatz.
\newblock Collective dynamics of `small-world'networks.
\newblock {\em Nature}, 393:440 EP --, 06 1998.

\end{thebibliography}

\end{document}